066

# Design and testing of low temperature part of an UHV – SPM microscope

Frolec J.[*], Vonka J.[*,**], Hanzelka P.[*], Kralik T.[*], Musilova V.[*], Urban P.[*]

[*] Institute of Scientific Instruments of the ASCR, v.v.i., 612 42 Brno, Czech Republic
[**] Faculty of Mechanical Engineering, Brno University of Technology, 616 69 Brno
Czech Republic

## ABSTRACT

We have developed low temperature part of an ultra high vacuum scanning probe microscope (UHV-SPM) working at variable temperature within the range from 20 K to 700 K. To achieve the required temperature range, a flow cooling system using cryogenic helium (~5 K) as a coolant was designed. The system consists of a flow cryostat and a flexible low-loss transfer line connecting a Dewar vessel with the flow cryostat. We have also tested liquid nitrogen (~ 77 K) as an alternative low-cost coolant. Using nitrogen, the microscope can operate at temperatures of about 100 K and higher. As the flow of the coolant through the cryostat can cause thermally induced two-phase flow fluctuations resulting in instability in temperatures, preliminary tests were done in order to find ways of avoiding the temperature oscillations and optimising the cooling process.

## 1.  INTRODUCTION

Our group has designed, built, and tested a low-temperature part of a new ultra-high vacuum scanning probe microscope (UHV–SPM) which will be built-in an UHV modular system based on scanning electron microscope (SEM). Major part of the design of the UHV-SPM cooling system is based on the work done in the frame of bachelor´s (Vonka 2011) and master´s (Vonka 2013) theses dealing with a similar UHV-STM cooling system. The complex UHV system with several other analytical methods will serve for *in situ* fabrication and characterization of nanostructures including surface analysis at atomic level. The UHV system with SEM and SPM is designed for study of samples at variable temperature within the range from 20 K to 700 K.

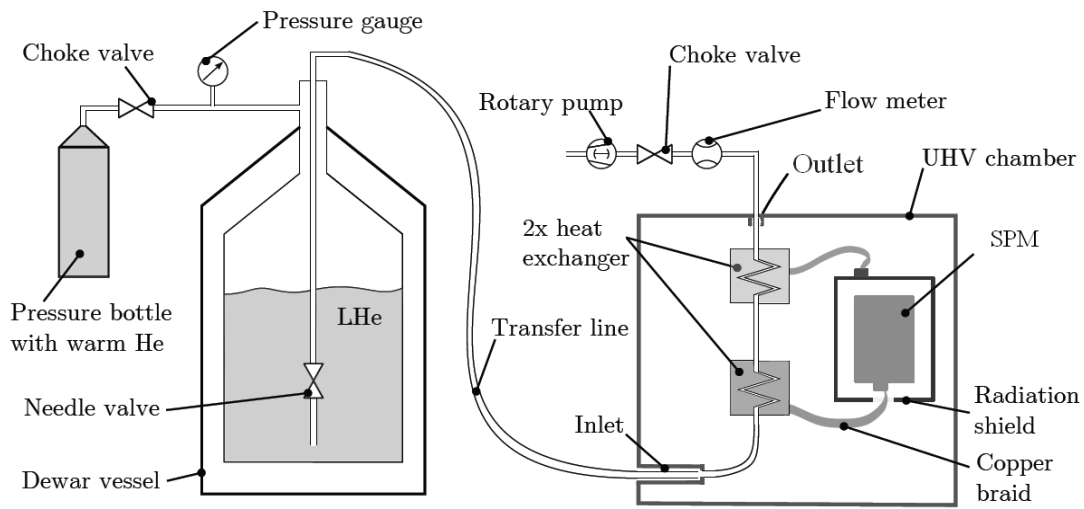

Figure 1.  The experimental setup scheme of low temperature cooling system of the UHV–SPM. Adapted from Vonka (2013).

Since a wide range of temperatures is required, a flow cooling system using liquid helium (LHe) or nitrogen (LN$_2$) as a coolant was developed. The cooling system is composed of a flow cryostat, a Dewar vessel and a flexible transfer line with vacuum insulation connecting both these parts together (Fig. 1 and 2). The flow cryostat is attached to an UHV chamber with the microscope and is assembled from an inlet and outlet part and two heat exchangers connected in series. The first colder heat exchanger is thermally linked via a copper braid to the sample holder of SPM while the second heat exchanger at higher temperature cools through another braid a radiation shield around the SPM. In this contribution, the design and tests of individual low temperature parts and of the whole flow cooling system are presented.

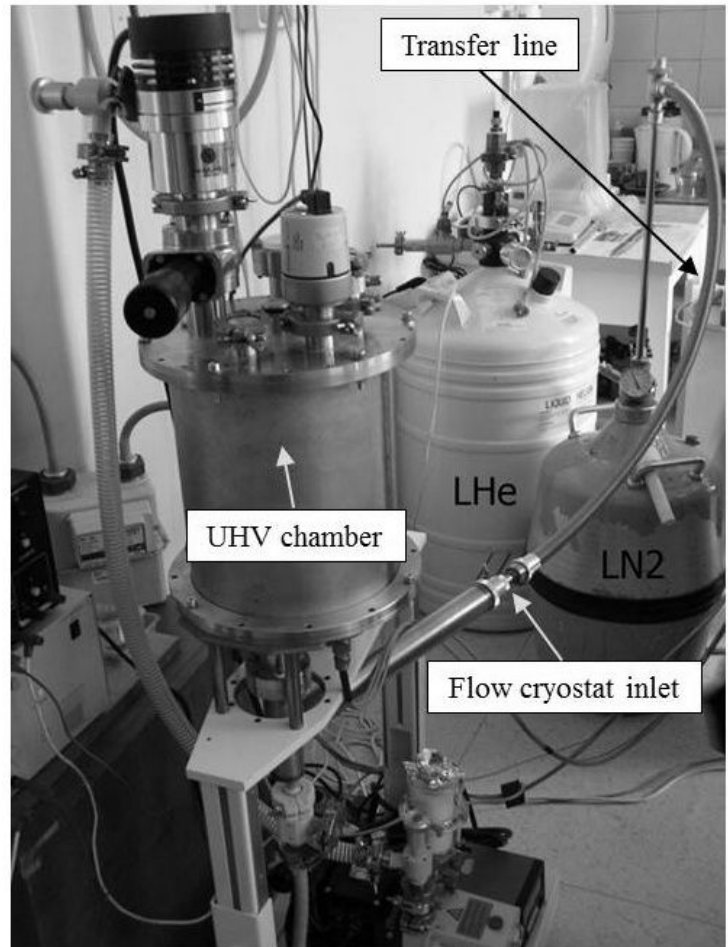

Figure 2. Experimental setup of the flow cooling system.

## 2. FLOW COOLING SYSTEM

The principle of the flow cooling system is as follows. The Dewar vessel is pressurized by warm He gas from external pressure bottle. Liquid helium (LHe) is then transferred via flexible transfer line with one end submersed in liquid He while the other end is inserted into an inlet tube of the flow cryostat (Fig. 2). The removable He low-loss transfer line has total parasitic heat loss of less than 900 mW. Its end is terminated with a connecting piece made of PTFE. Helium is then led from the transfer line to a T-shape assembly of the flow cryostat. Helium is transferred via a thin stainless steel capillary located inside an evacuated 400 mm long T-shape assembly made from stainless steel tube (40 mm in diameter). The T-shape assembly contains both the He inlet as well as outlet tubes.

In the vicinity of UHV–SPM there are two identical heat exchangers (Fig. 3 and 4) coupled in series and surrounded by a copper radiation shield thermally connected to the second heat exchanger. Each heat exchanger is made of copper cylinder with 300 mm long helical grove enclosed in a copper tube. The heat exchange takes place in the groove helix (with a diameter of 9.5 mm and cross-section area 0.75 $mm^2$), where He is forced to flow. Maximal calculated cooling power of the exchangers is 0.5 W (for 0.3 l/h of LHe and their temperatures 4.2 K and 20 K). Helium flows from the first to the second heat exchanger via 16 mm long thin-walled stainless steel tube. Each heat exchanger is equipped with a thermometer and resistive heater (500 Ω SMD resistor). The coolant temperature in the inlet capillary prior to entering the first heat exchanger is monitored as well.

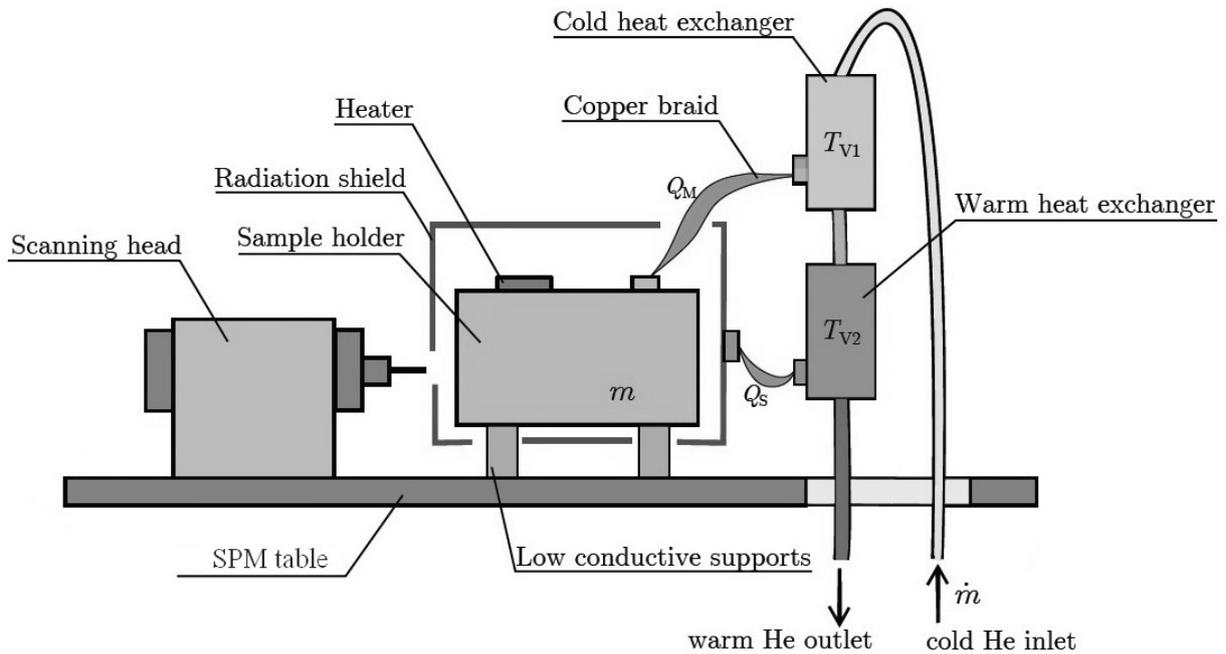

Figure 3. Low temperature part of UHV–SPM platform placed in vacuum chamber. Presented heat flows and temperatures are considered in calculations shown in Tabs. 1 and 2. The first heat exchanger (at temperature $T_{v1}$) cools down the sample holder in the UHV–SPM core, while the second one is thermally connected to the radiation shield. The SPM table is insulated from vibration by spring and electromagnetic damping. Adapted from Vonka (2013).

Thermal connection of heat exchangers with the sample holder and the radiation shield around the UHV–SPM is enabled by braids (Fig. 3). Each braid is 100 mm long and consists of three bundles containing 217 individual fine copper wires in each bundle with the single wire diameter of 75 μm (Vonka 2011). The braids are unravelled to individual wires in order to minimize the transfer of vibrations from the heat exchangers.

While the sample holder is cooled down to cryogenic temperature, the surrounding UHV chamber as well as the scanning probe stay at room temperature. For this reason, a suitable insulation with high mechanical stiffness and low thermal conductance is necessary. The thermal and mechanical criteria are fulfilled by using of three sets of four glass balls placed in vertices of a tetrahedron as a mounting of the sample holder (Hanzelka et al. 2013). We used balls made of glass with diameter of 2.5 mm. Low thermal conductivity of the four ball supports (FBS) is based on small contact areas between them. Such solution has the lowest thermal conductance among other feasible supporting structures. For example, calculated total heat flow through 10 mm long stainless steel tube (with wall thickness 0.1 mm) is more than ten times higher than in the case of FBS. Moreover, calculated heat flow through FBS was in a good agreement with the measured value (Hanzelka et al. 2013).

In order to test the cooling power of the cryostat and the thermal insulation ability of FBS, a model of the sample holder was created. This model is made of an aluminium alloy plate with dimensions of 30 x 30 x 4.5 mm and mass 26 g. It is equipped with a diode thermometer and 500 Ω SMD resistive heater. The model was placed on three FBSs inside the UHV chamber and thermally connected with the cold heat exchanger (Fig. 4). Final version of the sample holder will be probably made of CuCrZr alloy, which is characterised by high mechanical strength together with high and stable thermal conductivity (Hanzelka et al. 2010). Higher mass as well as higher heat capacity will also require higher cooling power (Vonka 2013).

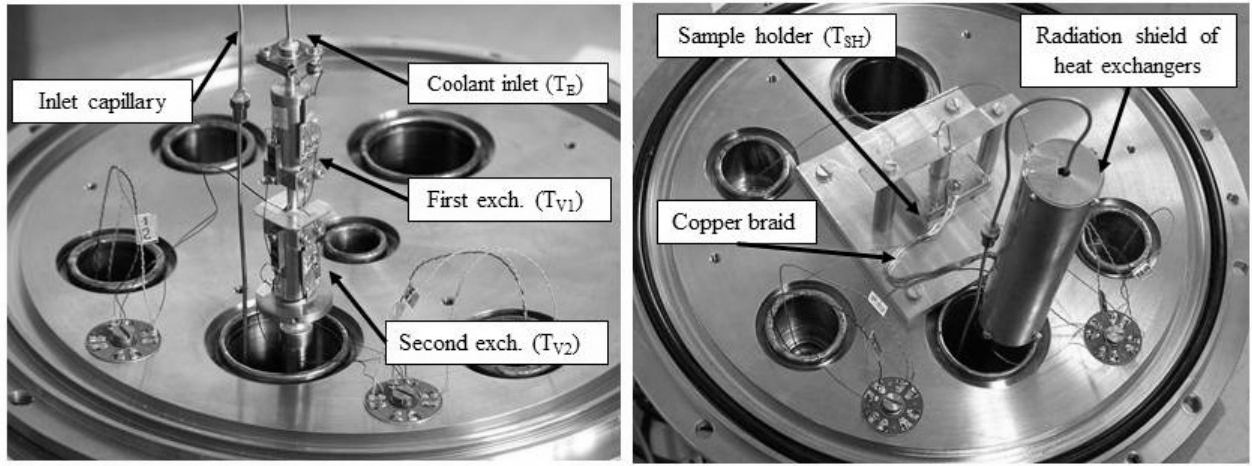

Figure 4. Experimental setup the heat exchangers (left) and sample holder (right) inside the test vacuum chamber.

## 3. HEAT FLOW ANALYSIS

From an economic point of view, the cooldown times as well as coolant consumption are very important features of any device operating at cryogenic temperatures and depend on the total heat load. Based on the heat flows calculation (Tab. 1 and 2), we have estimated the LHe consumption of the first (0.33 l/h) and the second (0.25 l/h) heat exchanger. Naturally, the LHe consumption for both of them must be in practice identical.

Table 1. Calculated heat flow $Q_M$ (see Fig. 3) to the first (cold) heat exchanger (at $T_{V1}$ = 10 K) together with estimation of corresponding LHe consumption. Adapted from Vonka (2013).

| Heat load on microscope | 260 mW |
|---|---|
| Radiation to the copper braid | 19 mW |
| Total heat load at the first heat exchanger | 279 mW |
| LHe consumption | 0.33 l/h |

Table 2. Calculated heat flow $Q_S$ (see Fig. 3) to the second (warm) heat exchanger (at $T_{V2}$ = 25 K) together with estimation of corresponding LHe consumption. Adapted from Vonka (2013).

| Heat load on the radiation shield | 580 mW |
|---|---|
| Radiation to the copper braid | 19 mW |
| Total heat load at the second heat exchanger | 599 mW |
| LHe consumption | 0.25 l/h |

Although the low temperature part of UHV–SPM was originally designed for liquid helium, performed tests have shown the possibility of using also liquid nitrogen ($LN_2$) as an alternative low-cost coolant. We have run several experiments using LHe as well as $LN_2$. Temperature controller Lakeshore LS-350 was used for temperature monitoring and regulation. It should be pointed out that we have tested the cooling system without the radiation shield, because the main goal was the testing of the flow cryostat system. Our experimental setup was able to cool down the sample holder model from room temperature to 20 K with the LHe consumption of 1.5 l/h. Much more economical is cooling to 25 K with helium flow rate 0.5 l/h, which takes about 50 min (Fig. 5a). The lowest achievable working temperature using $LN_2$ was about 100 K with the cooldown time of about 120 min and $LN_2$ consumption approximately 1.2 l/h (Fig. 5b). The consumption of the coolant needed for thermal stabilisation is generally lower and depends on selected temperature and system tuning. For instance, keeping sample holder at 25 K requires about 0.8 l/h of LHe, while the consumption for stabilisation at 50 K reaches about 0.32 l/h (Vonka 2011). In the case of $LN_2$ cooling, stabilisation at temperatures in temperature range 150 - 170 K took only 0.2 l/h. The temperature of the sample holder may be adjusted by the flow rate (using a needle valve at the end of transfer line in the Dewar vessel and a choke valve at the flow cryostat outlet) and by heating of the sample holder.

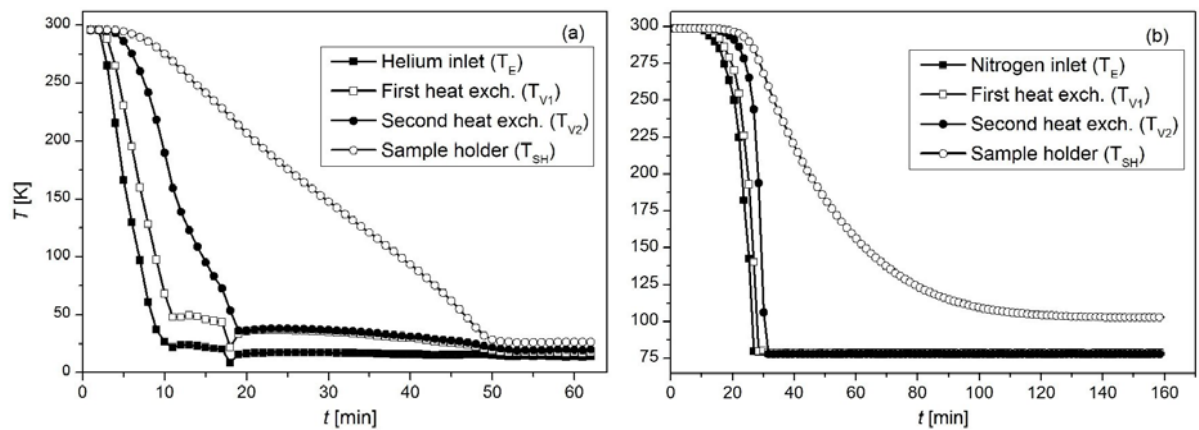

Figure 5. Time dependence of cooldown process using helium (a) and nitrogen (b). Consumption of cooling liquid is approximately 0.5 l (LHe) and 1.2 l (LN$_2$) per hour.

We have encountered some problems with temperature fluctuations caused by instable LHe flow. There was also a problem during LN$_2$ cooling with temperature stabilisation of heat exchangers at higher temperature setpoints. The cause may lie in not fully optimized cryostat setting or in partial evaporation of LHe / LN$_2$. The flow of the coolant through the cryostat can cause thermally induced two-phase flow oscillations, resulting in destabilisation of the cooling system (for a review see Kakac and Bon 2008). There are several ways to suppress these instabilities. For instance, it is possible to optimize the design of the system (Liang et al. 2011), select a tube with an optimal character of internal surface (Karsli et al. 2002), or even to avoid the two-phase flow regime completely by subcooling and pressurization of the coolant (Baek and Jeong 2012).

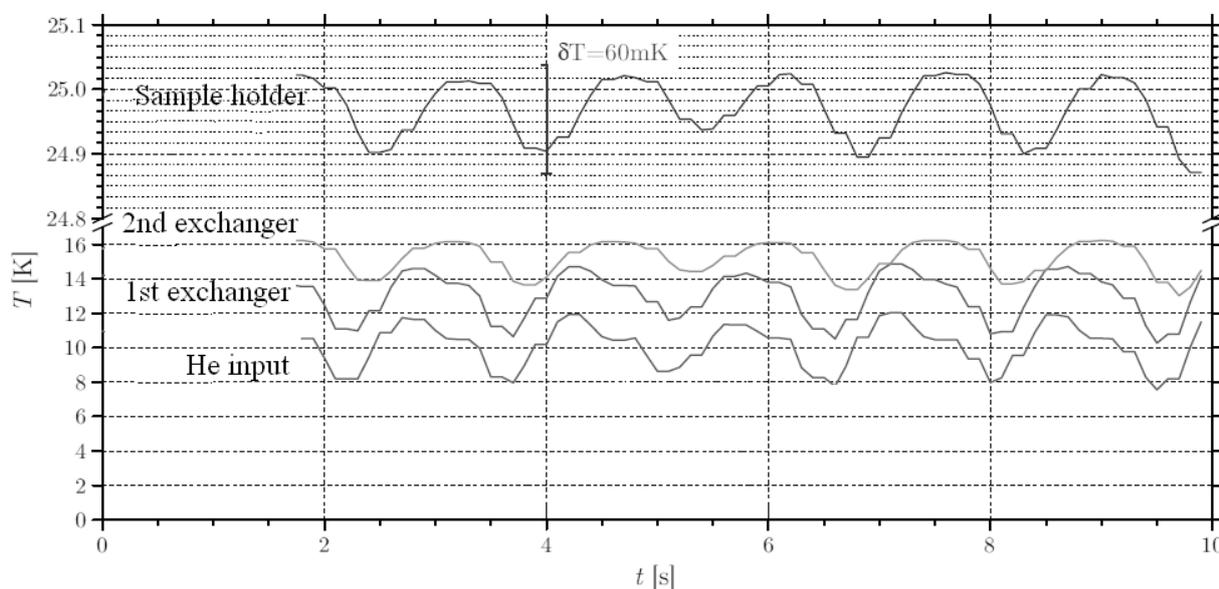

Figure 6. Temperature oscillations simultaneously detected during the process of cooling with LHe. Adapted from Vonka (2013).

In our case, we strive to optimize the stability of the system at selected temperature by adjusting the coolant flow rate and heating the sample holder. During the thermal stabilisation of the sample holder at 25 K, the amplitude of the temperature oscillations reached about 60 mK at sample holder and about 4.2 K at heat exchangers (Fig. 6). Increasing setpoint temperature resulted in decrease of oscillation amplitude. The reason why we use the liquid phase of the coolant is to reduce the cooldown time and to reach the lowest sample holder temperatures. Another reason of using liquid is to utilize the latent heat for cooling the flow cryostat, the most demanding heat-load part of our cooling system. However, for higher setpoint temperatures it may be possible to avoid two-phase flow oscillations by using only cold gas with lower cooling performance.

## 4. CONCLUSIONS

In conclusion, we have developed a flow cooling system for ultra-high vacuum scanning probe microscope. The main features of the assembly are versatility and low coolant consumption. Both of the commonly used cryogenic liquids, helium as well as nitrogen, can be used for cooling of the samples. The instrument will allow keeping the sample holder at desired temperature ranging from 20 K (100 K in the case of liquid nitrogen) up to 700 K.

## 5. ACKNOWLEDGMENTS

This work was supported by the Ministry of Education, Youth and Sports of the Czech Republic (LO1212) together with the European Commission (ALISI No. CZ.1.05/2.1.00/01.0017) and by TACR (Project No.TE01020233).